\documentclass{vgtc}                          

\graphicspath{{figures/}{pictures/}{images/}{./}} 
\usepackage{times}                     

\usepackage{booktabs}                  
\usepackage{xcolor}
\definecolor{kekecolor}{RGB}{180,40,120} 

\usepackage[most]{tcolorbox}
\definecolor{kekeblue}{RGB}{0,114,178}
\newtcolorbox{synthesisbox}{
  enhanced,
  breakable,
  frame hidden,
  colback=kekeblue!6,
  borderline west={3pt}{0pt}{kekeblue!70},
  arc=0mm,
  left=12pt,
  right=12pt,
  top=8pt,
  bottom=8pt,
  boxsep=0pt,
  before skip=10pt,
  after skip=10pt,
}
\usepackage{mathptmx}                  

\usepackage{hyperref}
\hypersetup{
  colorlinks=false,
  pdfborder={0 0 0}
}

\title{What Cognitive Accessibility Reveals About Data Visualization}

\author{
\makebox[\textwidth][c]{%
\begin{tabular}[t]{c}
Keke Wu\thanks{e-mail: kekewu@umd.edu}\\[-0.2em]
{\scriptsize University of Maryland, College Park}
\end{tabular}
\hspace{2em}
\begin{tabular}[t]{c}
Jinjuan Heidi Feng\thanks{e-mail: jfeng@towson.edu}\\[-0.2em]
{\scriptsize Towson University}
\end{tabular}
\hspace{2em}
\begin{tabular}[t]{c}
Jonathan Lazar\thanks{e-mail: jlazar@umd.edu}\\[-0.2em]
{\scriptsize University of Maryland, College Park}
\end{tabular}
}\\[0.3em]
{\scriptsize Maryland Initiative for Digital Accessibility}
}

\abstract{Data visualization aims to augment human cognition and make data accessible to diverse audiences. As data increasingly shapes participation and decision-making across many domains, there is a growing need to examine whether prevailing assumptions in visualization adequately reflect the diversity of human abilities, experiences, and needs. We argue that cognitive accessibility provides a critical lens for examining these questions and functions as a stress test for visualization theory. Drawing on cognitive accessibility research and our experiences studying accessible visualization, we identify three interconnected assumptions that shape visualization research and practice: assumptions about what forms of cognition visualization supports, how accessibility is defined and measured, and whose needs and abilities are centered in design and evaluation. Making these assumptions explicit reveals opportunities to rethink longstanding approaches and open new directions. Ultimately, we believe that cognitive accessibility can serve as a catalyst for innovation, expanding what visualization supports, whom it serves, and the roles it plays in people's lives.} 


\begin{document}
\maketitle
\section{Introduction}

Over the past decades, data visualization has transformed how people analyze, communicate, and make decisions with data. Foundational research has demonstrated how visual representations can amplify cognition and support sense-making~\cite{vision-to-think, cleveland1984graphical, bertin1983semiology}, while advances in visual encodings~\cite{ware2013information, wilkinson2005grammar}, interaction~\cite{Keim2002InformationVisualization, munzner2014visualization}, and evaluation~\cite{heer2010crowdsourcing, saraiya2005insight} have established visualization as a mature discipline with demonstrated impact across science, business, policy, education, and healthcare.

At the same time, visualization’s theories, methods, and evaluation practices reflect particular conceptions of cognition, accessibility, and audiences. Visualization research has traditionally emphasized analytical reasoning through representations optimized for accuracy, efficiency, and task performance~\cite{vision-to-think, ware2013information}, with evaluation similarly prioritizing correctness and efficiency. These choices have been instrumental in advancing the field, but they also reflect assumptions~\cite{shehryarreflection2026} about how people think with data, what counts as access, and whose needs and abilities are centered.

Cognitive accessibility research brings the limits of these assumptions into view. We approach cognitive accessibility as supporting diverse ways of understanding, interpreting, applying, and meaningfully engaging with information. Our perspective is grounded primarily in research with people with intellectual and developmental disabilities (IDD), including people with Down syndrome (DS). We situate this work within broader research on neurodivergence and cognitive disability, overlapping but distinct concepts that encompass diverse forms of cognitive functioning and experiences of disability. Across studies of visualization comprehension~\cite{wu2021dataaccessibility}, everyday data practices~\cite{wu2023datadataeverywhere}, and participatory design with people with IDD~\cite{wu2024codesign}, our research has examined how people understand, engage with, and represent data in different contexts. Studies with adults with DS~\cite{wood2024health} further highlight how visualization interpretation can depend on forms of literacy, reasoning, and cognitive resources often taken for granted. Together, these findings show how barriers to cognitive accessibility can expose implicit assumptions about how people understand and use data.

As visualization increasingly shapes participation and decision-making in everyday life, what cognitive accessibility reveals extends beyond any single population or accessibility challenge. Cognitive accessibility offers a lens for broadening conceptions of cognition, expanding accessibility beyond perception and usability, and reconsidering how people engage with and make meaning from data. What appear as isolated accessibility barriers can therefore reveal broader tensions between dominant models of visualization and the diverse ways people engage with data.

In this article, we examine what cognitive accessibility reveals about assumptions of cognition, accessibility, and audiences in visualization. Rather than attempting to characterize cognitive diversity comprehensively, we draw on our research with people with IDD and DS to identify and reflect on assumptions that may otherwise remain implicit. Making these assumptions visible, we argue, creates opportunities to rethink what visualization supports, whom it serves, and how its success is defined, and ultimately, to rethink visualization itself.
\section{Assumptions about Cognition}

Visualization research has traditionally been built around assumptions about how people attend to, reason with, and make sense of information. These assumptions often privilege abstract reasoning, sustained attention, working memory, and specific forms of prior knowledge as the primary pathways to understanding data. Cognitive accessibility research reveals that people draw on diverse reasoning strategies, attentional patterns, and forms of knowledge when engaging with data, challenging narrow conceptions of what it means to understand information.


\subsection{The Analytical Mind}
Many visualization techniques rely on abstract reasoning~\cite{munzner2014visualization}. Viewers are expected to interpret symbolic representations, identify relationships and patterns, connect graphical elements to underlying concepts, and draw conclusions that are not explicitly represented. Such tasks often require inference making, in which viewers move beyond what is directly visible to reason about underlying meanings, contexts, and implications.


Cognitive accessibility research reveals a broader range of strategies people use to make sense of information. Across our studies of graph comprehension, participatory design, and everyday data practices, understanding was often supported by familiar metaphors, meaningful contexts, and lived experiences rather than abstract representations alone~\cite{wu2021dataaccessibility, wu2023datadataeverywhere, wu2024codesign}. Participants frequently drew on personal experiences and existing knowledge to interpret data, while physical manipulation and embodied interaction helped make abstract concepts more tangible. Rather than serving only as supports for abstract reasoning, these strategies often formed the basis through which participants understood and interpreted data. Conversely, studies involving individuals with DS have found that challenges often emerge when graph comprehension requires abstract interpretation, ambiguity resolution, or multi-step inference making~\cite{wood2024health}.

These findings challenge the assumption that understanding relies primarily on abstraction and inference. Instead, cognitive accessibility highlights diverse reasoning practices, including metaphors, narratives, personal experience, affective engagement, and embodied interaction.


\subsection{The Attentive Mind}
Understanding a visualization is rarely a single-step process. It requires attending to relevant elements, integrating information across representations, and maintaining information in working memory while reasoning about data~\cite{baddeleyworkingmemory}. Visualization often assumes that viewers can sustain attention, manage cognitive load, and retain information across multiple stages of interpretation.


Cognitive accessibility research suggests that attention and memory vary substantially across audiences. People with cognitive disabilities often experience limitations in short-term memory that affect their ability to integrate and retain information~\cite{faragher2020}. Attention is also allocated differently: studies involving autistic individuals suggest that attention may be drawn to details that differ from those prioritized by conventional visual design~\cite{pelphrey2002}, while research with people with DS found that familiar elements such as text, icons, and colors often served as anchors during graph interpretation~\cite{wood2024health}. These differences influence what information is noticed, retained, and interpreted, suggesting that attention is not simply a prerequisite for visualization use but an active part of how meaning is constructed.

At the same time, cognitive accessibility research points to factors that can support attention and memory. Participants in studies of everyday data practices described how engaging formats such as games and films naturally guided their attention~\cite{wu2023datadataeverywhere}. Although individuals with DS may experience limitations in short-term memory, they demonstrate comparable long-term memory retention once information has been learned~\cite{faragher2020}, and emotional engagement may further support memory formation and retention~\cite{zimpel2016trisomy}.

These findings suggest that attention and memory are shaped by familiarity, engagement, and emotional connection as much as by cognitive capacity. Understanding therefore depends not only on what people can remember, but also on what captures their interest and remains meaningful over time. Cognitive accessibility reveals new possibilities for designing visualizations that actively guide attention and support memory.




\subsection{The Knowledgeable Mind}
Visualization is often described as offloading cognitive work onto visual perception. Yet interpreting visualizations is not automatic. It depends on multiple forms of prior knowledge, including print literacy, numeracy, and graphicacy. Cognitive disabilities can affect these abilities~\cite{Loveall2021reading, wood2024health}, while low literacy and numeracy are also common in the broader population~\cite{OECD, Mamedova_Pawlowski_2020}.

Of these literacies, graphicacy is particularly important for interpreting visualization. Curcio~\cite{curcio1987comprehension} describes graph comprehension as progressing from reading data, to reading between data, to reading beyond data. Among adults with DS, performance was relatively strong when identifying visual elements and values, but declined as tasks required comparison, inference, and interpretation~\cite{wood2024health}. Similar challenges emerged in graphical perception and co-design studies, where conventional visualizations, especially pie charts~\cite{wu2021dataaccessibility}, could become barriers to understanding, prompting participants to draw on concrete analogies and familiar experiences when interpreting visual representations~\cite{wu2024codesign}.

Together, these findings suggest that visualization assumes a shared visual language grounded in multiple forms of literacy. Cognitive accessibility reminds us that this knowledge cannot be taken for granted across audiences, challenging visualization researchers to reconsider the literacy demands embedded in conventional representations.

\begin{synthesisbox}
\textbf{Rethinking Cognition.}
Cognitive accessibility suggests that visualization should broaden what it recognizes as legitimate pathways to understanding. Rather than optimizing primarily for analytical reasoning, visualization can better support diverse ways of reasoning, attending to, and making sense of data through emotion, embodiment, and lived experience.
\end{synthesisbox}

\section{Assumptions about Accessibility}

Access is a central goal of data visualization, but it is often understood through three related perspectives: perception, performance, and accommodation. While these perspectives have expanded accessibility research, they capture only part of what it means to access data. Cognitive accessibility broadens access beyond seeing information, completing tasks, or adapting existing designs to include how people interpret, apply, and engage with data in real life.

\subsection{Access as Perception}
Visualization accessibility has traditionally focused on perception. Researchers have developed perceptually effective visual encodings~\cite{cleveland1984graphical, ware2013information}, color-accessible designs~\cite{brewer2003}, and alternative modalities that reduce perceptual barriers and improve access to visual information~\cite{lundgard2022natural, elvasky2024navigator, hoque2023naturalsound, marriott2026vision}. These efforts have been instrumental in expanding visualization access.

However, this perspective often treats accessibility as a problem of perception and decoding. Cognitive accessibility research suggests that perceptual access is necessary but not sufficient for data use. Research with DS has shown that barriers frequently emerge not from seeing information, but from interpreting it, relating it to everyday experiences, or applying it in meaningful ways~\cite{wood2024health}. People may successfully identify visual elements, values, and patterns while still struggling to understand what the data means or how it is relevant. These findings suggest that accessibility extends beyond making information perceptible to supporting interpretation, contextualization, and meaningful use.

\subsection{Access as Performance}
Measures such as task accuracy, completion time, and error rates are widely used to evaluate visualization effectiveness and comprehension. Rooted in early HCI (Human-Computer Interaction) and fields such as human factors and psychology~\cite{lazar2017research}, these standardized metrics assume that success on analytical tasks reflects successful access. As a result, accessibility is often evaluated through performance on well-defined tasks.

While these measures provide evidence of usability, they privilege a particular kind of interaction with data: extracting information and answering questions. Researchers have argued that performance on laboratory tasks does not necessarily capture broader outcomes that matter in real-world settings, such as motivation, collaboration, trust, empathy, and social participation~\cite{lazar2017research, shneiderman2011}. Cognitive accessibility similarly challenges the assumption that analytical performance is the primary way people engage with data. In everyday life~\cite{wu2023datadataeverywhere}, people also use data to reflect on experiences, express themselves, advocate for their needs, manage health, and support everyday autonomy.

Similar forms of engagement emerged in participatory design and co-design research. Participants used data to express feelings, communicate identity, and tell stories about their lives~\cite{wu2024codesign}. Reflecting on these representations often fostered new insights, confidence, understanding, and empowerment.

These findings suggest that performance on analytical tasks provides only a partial account of access. Cognitive accessibility broadens this view by recognizing forms of engagement in which visualization supports reflection, self-expression, emotional well-being, and everyday life.

\subsection{Access as Accommodation}
Accessibility is often approached as a process of adapting visualizations to meet the needs of particular audiences. This perspective has motivated important work on accessible encodings~\cite{khalaila2026tactile}, assistive technologies~\cite{zongrich2022, thompson2023chartreader}, and alternative representations~\cite{hoque2023naturalsound, marriott2026vision} that reduce barriers for people with disabilities. Such efforts have expanded who can access and engage with data.

However, accessibility is often addressed only after key decisions about representation, interaction, and evaluation have already been made. In these cases, it becomes a retrofit. While retrofitting can reduce barriers, it is often costly and remains constrained by assumptions embedded in the original design~\cite{lazar2026born}. The focus shifts to making existing visualizations more accessible, rather than questioning whether those representations and interactions are appropriate in the first place or exploring alternative ways of engaging with the underlying data.

Cognitive accessibility research suggests that involving people with cognitive disabilities early can reveal opportunities that are difficult to identify through adaptation alone. Participatory and co-design studies have produced representations and forms of engagement grounded in participants’ lived experiences, interests, and communication practices~\cite{wu2024codesign}. In these cases, accessibility did not simply improve existing visualizations; it shaped what was designed.

These findings suggest that accessibility is not only a design constraint but also a source of design insight. Rather than retrofitting existing visualizations, cognitive accessibility points toward \textit{born-accessible visualization}~\cite{lazar2026born}, where diverse ways of thinking, communicating, and engaging with data shape design from the outset.

\begin{synthesisbox}
\textbf{Rethinking Accessibility.}
Cognitive accessibility reframes accessibility as more than removing barriers to existing visualizations. Instead, it positions accessibility as supporting interpretation, participation, and meaningful engagement, broadening both what visualization is designed to support and how its success is evaluated.
\end{synthesisbox}
\section{Assumptions about Audiences}

Conventional visualization often assumes that audiences seek utility from data, are ready to engage with it, and share sufficiently similar needs to be supported through generalizable design principles. Cognitive accessibility research reveals substantial variation in the value people seek from data, their relationships with it, and the forms of support that enable meaningful engagement.

\subsection{The Utilitarian Audience}
Visualization research often frames data engagement in terms of utility: extracting patterns, understanding information, communicating insights, and supporting decisions. Under this view, the value of visualization is largely determined by how effectively it helps people achieve these goals.

Research with people with IDD suggests that data engagement can also be valuable as an experience in itself~\cite{wu2023datadataeverywhere, wu2024codesign}. In these contexts, the goal is not always to solve a problem or reach a decision, but to explore, reflect, and connect with one’s experiences. Participants described enjoyment in creating data representations, curiosity about exploring their own lives, and satisfaction in discovering new patterns about themselves. Data engagement was often playful, creative, and personally meaningful, providing opportunities for exploration, expression, and connection without a predefined practical goal.

These findings reveal the limits of a narrowly utilitarian view of visualization, where value is defined primarily by task efficiency and performance. Cognitive accessibility recognizes experiential forms of value, where curiosity, creativity, enjoyment, and personal meaning are worthwhile outcomes in their own right.

\subsection{The Data-Ready Audience}
Visualization often assumes audiences are already prepared to engage with data. Viewers are expected to recognize data as relevant, see themselves as legitimate data users, and willingly participate in data-driven activities. As a result, visualization systems typically focus on helping people interpret and act on information rather than fostering engagement itself.

Many people, however, do not begin from this position~\cite{wu2023datadataeverywhere}. Participants with disabilities often described data as technical, difficult, or remote, and many did not see themselves as people who work with data. Yet many were already collecting information, tracking experiences, sharing observations, and reflecting on everyday life in ways that closely resembled data practices. What was often missing was recognizing these activities as forms of data engagement. Engagement emerged when data became connected to familiar experiences, personal goals, and everyday concerns.

These observations challenge the assumption that audiences arrive ready to engage with data. Engagement depends not only on access to information, but also on whether data feels relevant to people’s experiences, interests, and goals. Cognitive accessibility highlights the importance of helping people see themselves in relation to data, suggesting that fostering identification with data may be as important as supporting interpretation.

\subsection{The Universal Audience}
Visualization research often seeks design principles that apply broadly across audiences. This pursuit has produced foundational knowledge about visual encoding, interaction design, and graphical perception, enabling the development of widely applicable visualization techniques.

Research with cognitively diverse and disabled audiences reveals important variation in how people interact with data. Individuals with DS often benefited from concrete representations grounded in familiar experiences, people with cerebral palsy often required accommodations for co-occurring vision and mobility impairments, while autistic participants often preferred simpler, less visually cluttered visualizations that reduced sensory demands~\cite{wu2021dataaccessibility, wood2024health}. Participatory design studies further showed that personal interests and emotions shaped both what participants attended to and how they chose to represent data~\cite{wu2024codesign}. Together, these findings suggest that no single design approach is equally effective across audiences or contexts.

The idea of a universal audience therefore has limits~\cite{shehryarreflection2026}. Broad design principles remain valuable, but they cannot fully account for the diversity of how people think, feel, and engage with data. Rather than treating this variation as deviations from a universal user, cognitive accessibility treats it as a source of design insight, opening new opportunities for creating richer and more flexible visualization experiences.

\begin{synthesisbox}
\textbf{Rethinking Audiences.}
Cognitive accessibility calls into question the notion of a universal, data-ready audience with shared goals and motivations. Rather than designing for a single ``typical'' user, visualization should embrace diverse relationships to data shaped by people’s experiences, interests, and identities.
\end{synthesisbox}
\section{Toward Cognitive Diversity in Visualization}

Throughout this paper, we have argued that cognitive accessibility functions as a productive stress test for visualization theory. Examining visualization through this lens reveals assumptions about cognition, accessibility, and audiences that often remain invisible when designing for cognitively normative users. In doing so, it exposes both the limits of these assumptions and alternative ways people understand, engage with, and derive value from data.

This perspective challenges several longstanding views in visualization. Understanding may emerge through embodied, emotional, and experience-based forms of engagement, not only abstraction and analytical reasoning. Accessibility extends beyond perception and task performance to include interpretation, participation, and meaningful use. Audiences differ not only in ability, but also in their motivations, experiences, and relationships to data.

Rather than treating cognitive diversity as an edge case, we argue that it offers a lens for reexamining the foundations of visualization. Attending to cognitive diversity can help develop theories, methods, and designs that better reflect the diversity of human experience, expanding what visualization can support, who it can serve, and what meaningful engagement with data can become.


\bibliographystyle{abbrv/abbrv-doi-hyperref}

\bibliography{main}
\end{document}